\documentstyle[twocolumn,epsf]{jpsj}
\newcommand{\simg}{\stackrel{>}{_\sim}}
\newcommand{\siml}{\stackrel{<}{_\sim}}

\def\runtitle{Critical Behaviour near the Mott Metal-Insulator Transition in a Two-band Hubbard Model
}
\def\runauthor{Yasuo {\sc{Ohashi}} and Yoshiaki {\sc{\=Ono}}}

\title
{
Critical Behaviour near the Mott Metal-Insulator Transition in a Two-band Hubbard Model
}

\author
{Yasuo {\sc{Ohashi}}* and Yoshiaki {\sc{\=Ono}}**}

{}

\recdate
{
}

\abst
{
The Mott metal-insulator transition in the two-band Hubbard model in infinite dimensions is studied by using the linearized dynamical mean-field theory. The discontinuity in the chemical potential for the change from hole to electron doping is calculated analytically as a function of the on-site Coulomb interaction $U$ at the $d$-orbital and the charge-transfer energy $\Delta$ between the $d$- and $p$-orbitals. Critical behaviour of the quasiparticle weight is also obtained analytically as a function of $U$ and $\Delta$. The analytic results are in good agreement with the numerical results of the exact diagonalization method.
}

\kword
{Mott transition, metal-insulator transition, two-band Hubbard model, dynamical mean-field theory, infinite dimensions}

\def\nn{\nonumber}

\begin{document}
\sloppy
\maketitle
\footnotetext{* E-mail: ohashi@edu2.phys.nagoya-u.ac.jp}
\footnotetext{** E-mail: c42545a@nucc.cc.nagoya-u.ac.jp}

\section{Introduction}
\label{sec:Intro}

The Mott metal-insulator transition (MIT) driven by the electron correlation is a fundamental problem in the condensed matter physics. Recently, some significant progress has been achieved in understanding the MIT by using the dynamical mean-field theory (DMFT) \cite{Georges1} which becomes exact in the limit of infinite spatial dimensions \cite{Metzner}. In this approach, the lattice problem is mapped onto an effective impurity problem where a correlated impurity site is embedded in an effective uncorrelated medium that has to be determined self-consistently. 
To solve the effective impurity problem, several methods have been applied including the iterated perturbation theory \cite{Georges1}, the non-crossing approximation \cite{Pruschke}, the projective self-consistent method (PSCM) \cite{Fisher}, the quantum Monte Carlo (QMC) method \cite{Jarrell}, the exact diagonalization (ED) method \cite{Caffarel} and the numerical renormalization group (NRG) method \cite{Sakai,Bulla1}.

In the single-band Hubbard model on the infinite dimensional Bethe lattice, the Mott MIT is found to occur as a first-order phase transition at finite temperature below a critical temperature $T_c$ \cite{Georges1}. Below $T_c$, a coexistence of the metallic and insulating solutions is found for the same value of the on-site Coulomb interaction $U$ in the range $U_{c1}(T)<U<U_{c2}(T)$ \cite{Georges1,Rozenberg,Schlipf,Blumer,Krauth,Joo,Bulla4}. At zero temperature, coexistence is also obtained for $U_{c1}<U<U_{c2}$ \cite{Caffarel,Bulla1}. 
When $U$ increases below $U_{c2}$, the quasiparticle weight in the metallic solution decreases and finally becomes zero in the limit $U \to U_{c2}$. When $U$ decreases above $U_{c1}$, the energy gap in the insulating solution decreases and finally becomes zero in the limit $U \to U_{c1}$. The ground state energy in the metallic solution is lower than that in the insulating solution for $U_{c1}<U<U_{c2}$. Therefore the Mott MIT occurs at $U=U_{c2}$ as a continuous transition at $T=0$. In this paper we will concentrate on the Mott MIT at $T=0$ and, then, we will denote the critical value $U_{c2}$ simply by $U_c$.

The Mott MIT is observed in various $3d$ transition-metal compounds, which are classified into two types: the Mott-Hubbard (MH) type and the charge-transfer (CT) type \cite{Fujimori,Zaanen}. In the MH type such as Ti and V compounds, the Coulomb interaction $U$ at the $d$-orbital is smaller than the CT energy $\Delta$ between $d$- and anion $p$-orbitals. In this case, the energy gap of the insulator is given roughly by $U$ and a MIT occurs at a critical value $U_c$ when $U$ is varied. In the CT type such as Co, Ni and Cu compounds, $U$ is larger than $\Delta$. Then the energy gap is roughly given by $\Delta$ and a MIT occurs at a critical value $\Delta_c$ when $\Delta$ is varied. 

In this paper, we investigate the Mott MIT with both the MH type and the CT type. We, therefore, need to use the two-band Hubbard model which is characterized by two parameters: the on-site Coulomb interaction $U$ at the $d$-orbital and the CT energy $\Delta$ between the $d$- and $p$-orbitals. Several authors have studied the model using the DMFT approach \cite{Caffarel,Georges2,Mutou,Ono2,Ono1}.  However, numerical problems make it difficult to study the Mott MIT for this model, in contrast to the single-band Hubbard model. In the half-filled single-band Hubbard model on the Bethe lattice, the chemical potential is fixed to $\mu=\frac{U}{2}$ because of the particle-hole symmetry. On the other hand, for the two-band Hubbard model, the Mott MIT occurs away from particle-hole symmetry and, then, the chemical potential has to be determined explicitly to fix the electron density per unit cell to be unity. This calculation consumes a lot of CPU time.  Furthermore, the Mott MIT point is a function of several  parameters, all of which have to be calculated, making it a difficult numerical problem.

Recently, a linearized form of the DMFT has been developed, where the dynamical mean-field equation is linearized near the Mott MIT \cite{Bulla,Ono,Ono3}. The linearized DMFT provides a simple and attractive technique to obtain approximate but analytical results for the critical regime of the MIT. In the single-band Hubbard model, it has been demonstrated that the critical interactions of the MIT predicted by this approach is in very good agreement with most accurate numerical estimates \cite{Bulla}. The linearized DMFT has been extended to the particle-hole asymmetric case. This allows for a comprehensive analytical investigation of the critical behaviour as a function of the on-site Coulomb interaction and the doping \cite{Ono3}. The discontinuity in the chemical potential on changing from hole to electron doping has been calculated analytically and is found to be in good agreement with the results of numerical methods, NRG as well as ED. Furthermore, analytic expressions for the compressibility, the quasiparticle weight, the double occupancy and the local spin susceptibility near half-filling, have been derived as functions of the interaction and the doping. These are difficult to calculate using the numerical methods mentioned above. The linearized DMFT has also been extended to the two-band Hubbard model \cite{Ono}. The phase boundary of the MIT is obtained analytically over the whole parameter regime including the MH type and the CT type. The analytical result agrees well with the numerical result obtained from the ED method. However, the critical behaviour of the two-band Hubbard model was not considered there. 

In the present paper, we study the critical behaviour near the Mott MIT in the two-band Hubbard model by using the linearized DMFT. The model and the formulation are detailed in \S 2. In \S 3, we analytically calculate the discontinuity in the chemical potential for the change from hole to electron doping as a function of $U$ and $\Delta$ over the hole parameter regime including the MH type and the CT type. The analytical results are in good agreement with the numerical results calculated by using the ED method. In \S 4, we also analytically obtain the quasiparticle weight near the MIT as a function of $U$ and $\Delta$. The analytic results are compared with the ED results for several parameters. Finally, discussions are given in \S 5.

\section{Linearized Dynamical Mean-Field Theory}
\label{sec:MIT}

\subsection{Single-band Hubbard model}
\label{sec:1band}

First, we consider the single-band Hubbard model, 
\begin{eqnarray} 
H= - \sum_{<i,j>,\sigma} t_{i,j} (c_{i\sigma}^{\dagger} c_{j\sigma} 
      + h.c.) + U \sum_{i} n_{i\uparrow}n_{i\downarrow}.  \label{HUB}
\end{eqnarray} 
In the limit of infinite dimensions, the self-energy becomes purely site-diagonal and the local Green's function $G(z)$ can be given by the impurity Green's function of an effective single impurity Anderson model, 
\begin{eqnarray} 
H_{\rm And}&=& \varepsilon_f \sum_{\sigma} f^{\dagger}_\sigma f_\sigma + U f^{\dagger}_{\uparrow}f_{\uparrow} f^{\dagger}_{\downarrow}f_{\downarrow} \nonumber \\ 
&+&\sum_{k,\sigma} \varepsilon_k c^{\dagger}_{k\sigma} c_{k\sigma} +\sum_{k,\sigma}V_k(f^{\dagger}_{\sigma}c_{k\sigma}+c^{\dagger}_{k\sigma}f_{\sigma}), 
\label{AND}
\end{eqnarray} 
where $\varepsilon_f$ is the impurity level and $\varepsilon_k$ are energies of conduction electrons hybridized with the impurity by $V_k$. 
In the model eq. (\ref{AND}), the non-interacting impurity Green's function, 
\begin{eqnarray} 
{\cal G}_0(z) = (z-\varepsilon_f- \Delta(z))^{-1}, \label{G0}
\end{eqnarray} 
with the hybridization function, 
\begin{eqnarray} 
\Delta(z)=\sum_k \frac{V_k^2}{z-\varepsilon_k}, \label{DELTA}
\end{eqnarray} 
includes effects of the interaction at all the sites except the impurity site and is determined self-consistently so as to satisfy the self-consistency equation.

For simplicity, the calculations in this paper are restricted to the Bethe lattice with the connectivity $q$ and the hopping $t_{i,j}=\frac{t}{\sqrt{q}}$. In the limit $q=\infty$, the self-consistency equation is given by 
\begin{equation} 
{\cal G}_0(z)^{-1} = z+\mu-t^2 G(z), \label{SCE}
\end{equation} 
where $\mu$ is the chemical potential for the original lattice model. 
Because of the particle-hole symmetry at half-filling, the chemical potential and the impurity level are set to $\mu = \frac{U}{2}$ and $\varepsilon_f = -\frac{U}{2}$, respectively. Then, the self-consistency equation (\ref{SCE}) is simply written by 
\begin{eqnarray} 
\Delta(z)=t^2 G(z). \label{SCE2}
\end{eqnarray}

The effective impurity problem  have been solved by using various numerical methods \cite{Georges1}. In the metallic phase with intermediate interaction, the density of states is characterized by a three-peak structure consisting of the upper and the lower Hubbard bands and a quasiparticle peak near the Fermi level. When the system approaches the MIT from the metallic side at $T=0$, the quasiparticle peak is found to be isolated from the upper and the lower Hubbard bands \cite{Bulla1}. As $U$ increases the width of the quasiparticle peak becomes narrow and finally vanishes in the limit $U\to U_c$. 

The linearized DMFT \cite{Bulla,Ono,Ono3} focuses on this critical regime close to the MIT, where the quasiparticle peak is approximated by a single pole at the Fermi level, i.\ e.\, 
$
G(z) = \frac{Z}{z},
$ 
near the Fermi level with a small quasiparticle weight $Z \to 0$ as $U \to U_c$. 
Correspondingly, the hybridization function $\Delta(z)$ is a single-pole function, 
\begin{equation}
\Delta(z)=\frac{V^2}{z} \label{DELTA0}. 
\end{equation}
This represents an approximate mapping of the model (\ref{HUB}) onto a two-site  Anderson model \cite{Hewson}, 
\begin{eqnarray} 
H_{\rm 2-site}&=& \epsilon_f \sum_{\sigma} f^{\dagger}_\sigma f_\sigma +U f^{\dagger}_{\uparrow}f_{\uparrow}f^{\dagger}_{\downarrow}f_{\downarrow} \nonumber \\ 
&+&\epsilon_c \sum_{\sigma} c^{\dagger}_{\sigma} c_{\sigma} + V \sum_{\sigma}(f^{\dagger}_{\sigma}c_{\sigma}+c^{\dagger}_{\sigma}f_{\sigma}), 
\label{2SITE}
\end{eqnarray} 
with $\epsilon_c=0$ and $\epsilon_f=-\mu$. The hybridization strength $V$ has be determined from the self-consistency equation which takes the simple form
\begin{eqnarray} 
t^2 Z = V^2. \label{SCE3}
\end{eqnarray}

Now we calculate the critical value of $U$ for the MIT at half-filling. In this case, the chemical potential is fixed to $\mu=\frac{U}{2}$ because of the particle-hole symmetry. Then the quasiparticle weight $Z$ is given by \cite{Hewson} (see also Appendix~\ref{appendix1})
\begin{eqnarray} 
Z = 36 \frac{V^2}{U^2}, \label{W}
\end{eqnarray} 
up to second order in $V$. From eqs. (\ref{SCE3}) and (\ref{W}), we obtain the critical value of the MIT in the linearized DMFT
\begin{eqnarray} 
U_c = 6t. \label{UC}
\end{eqnarray}
The analytic result eq.(\ref{UC}) is in good agreement with the best numerical estimates of NRG ($U_c=5.88t$) \cite{Bulla1} and ED ($U_c=5.87t$) \cite{Ono} as well as PSCM ($U_c=5.84t$) \cite{Georges1}. The iterated perturbation theory as well as the Gutzwiller approximation gives a larger critical value $U_c=6.6t$ \cite{Georges1} as compared to the numerical approaches mentioned above.

When we solve the self-consistency equation (\ref{SCE3}) with eq.(\ref{W}) by iteration, $V^2$ increases exponentially with iteration number for $U<U_c$ and, then, the single pole approximation for $\Delta(z)$  breaks down, resulting in the metallic solution. For $U>U_c$, $V^2$ decreases exponentially to give the self-consistent value $V^2 = 0$ corresponding to the insulating solution. 

For $U<U_c$, the chemical potential $\mu$ is continuous at $n=1$ as a function of $n$. On the other hand, for $U>U_c$, $\mu$ has a discontinuity at $n=1$. To calculate the discontinuity within the linearized DMFT, we need the result of the quasiparticle weight $Z$ in the general case including non-symmetric case with $\mu \ne \frac{U}{2}$ \cite{Ono3} (see also Appendix~\ref{appendix1}),
\begin{eqnarray} 
Z=F(U,\mu)V^2, 
\end{eqnarray} 
where 
\begin{eqnarray} 
F(U,\mu)=\frac{5}{2\mu^2}+\frac{4}{\mu(U-\mu)}+\frac{5}{2(U-\mu)^2}. 
\label{F} 
\end{eqnarray} 
When $U=U_{c}$, $F(U,\mu)$ has a minimum at $\mu=\frac{U}{2}$ and $t^2F(U,\frac{U}{2})=1$ which yields the critical value of the MIT given in eq.(\ref{UC}). When $U<U_{c}$, $t^2F(U,\mu)>1$ for all $\mu$, and the system is metallic for all $n$. When $U>U_{c}$, $t^2F(U,\mu)<1$ for $\mu_- <\mu< \mu_+$ resulting in the insulating state at $n=1$, while, $t^2F(U,\mu)>1$ for $\mu<\mu_-$ or $\mu>\mu_+$ resulting in the metallic state at $n\ne 1$. Then the chemical potential shows a discontinuity ${\it \Delta}\mu=\mu_+-\mu_-$ at $n=1$ for the change from electron to hole doping, where $\mu_\pm$ is given by the equation 
\begin{eqnarray} 
t^2F(U,\mu_\pm)=1, \label{MUC}
\end{eqnarray} 
for $U>U_c$. 
By solving eq.(\ref{MUC}) with eq.(\ref{F}), we obtain the discontinuity in the chemical potential 
\begin{eqnarray} 
{\it \Delta}\mu= U\left(1+\frac{1}{18u^2}-\sqrt{\frac{10}{9u^2}
        +\left(\frac{1}{18u^2}\right)^2} \right)^{\frac{1}{2}}, \label{Delmu}
\end{eqnarray}  
where $u \equiv \frac{U}{U_{c}}>1$. For $U>U_{c}$ close to $U_{c}$, eq.(\ref{Delmu}) yields 
$
{\it \Delta}\mu=\frac{6}{\sqrt{38}}U_{c}\sqrt{\frac{U}{U_{c}}-1}, 
$
while, for $U\gg U_c$, it yields ${\it \Delta}\mu\sim U$. 
The analytic result eq.(\ref{Delmu}) agrees well with the numerical results of NRG and ED as well as PSCM \cite{Ono3}. 
We note that the discontinuity ${\it \Delta}\mu$ calculated within the metallic solution is different (smaller) compared with the energy gap within the insulating solution \cite{Fisher}. In fact, ${\it \Delta}\mu$ is zero for $U\leq U_c=U_{c2}$, while the energy gap is finite for $U>U_{c1}$, where $U_{c1}<U_{c2}$ as mentioned in \S 1.

Finally, we study the critical behaviour near the Mott MIT \cite{Bulla,Ono3}. In this case, we need the result of the quasiparticle weight $Z$ up to fourth order in $V$ , which is given by \cite{Bulla} (see also Appendix~\ref{appendix1}),
\begin{eqnarray} 
Z = 36 \frac{V^2}{U^2}\left(1-44\frac{V^2}{U^2}\right), \label{W2}
\end{eqnarray} 
in the particle-hole symmetric case with $\mu=\frac{U}{2}$ \cite{critical}. From eqs. (\ref{SCE3}) and (\ref{W2}), we obtain the quasiparticle weight 
\begin{eqnarray} 
Z = \frac{18}{11}\left(1-\frac{U}{U_{c}} \right), \label{Z2} 
\end{eqnarray} 
near $U_c$ for $U<U_c$. The critical property eq.(\ref{Z2}) near $U_c$ obtained from the linearized DMFT is similar to that from the Gutzwiller approximation which predicts \cite{Vollhardt} 
$
Z = (1-\frac{U}{U_{c}} )^2. 
$
However, the coefficients $C$ of the quasiparticle weight 
$
Z = C (1-\frac{U}{U_{c}}) 
$
near $U_c$ are different, which are $C=2$ within the Gutzwiller approximation and $C=18/11$ within the linearized DMFT. The analytic result $C=18/11$ seems to be still too large compared with numerical results of NRG ($C \ll 1$) \cite{Bulla1} and ED ($C \approx 0.3$) \cite{Ono} as well as PSCM ($C \approx 0.9$) \cite{Georges1}. But a precise value of $C$ is not obtained for the present.

\subsection{Two-band Hubbard model}
\label{sec:2band}

The linearized DMFT has been extended to the two-band Hubbard model \cite{Ono}, 
\begin{eqnarray} 
H &=& \frac{t_{pd}}{\sqrt{q}} \sum_{<i,j>,\sigma}
        ( d_{i\sigma}^{\dagger} p_{j\sigma} + h.c.)
      + U \sum_{i} d^{\dagger}_{i\uparrow}d_{i\uparrow} 
                   d^{\dagger}_{i\downarrow}d_{i\downarrow}  
        \nonumber \\
  & & {}+  \epsilon_{d0} \sum_{i,\sigma} d_{i\sigma}^{\dagger} d_{i\sigma} + 
  \epsilon_{p0} \sum_{j,\sigma} p_{j\sigma}^{\dagger} p_{j\sigma}.         
\label{DP}
\end{eqnarray} 
This model eq.(\ref{DP}) is characterized by three parameters: the hopping integral $t_{pd}$ between the $d$- and $p$-orbitals, the on-site Coulomb interaction $U$ at the $d$-orbital and the charge-transfer energy $\Delta=\epsilon_{p0}-\epsilon_{d0}$ between the $d$- and $p$-orbitals. Henceforth we set $t_{pd}=1$, unit of energy, and $\epsilon_{d0}=0$, origin of energy, then, $\epsilon_{p0}=\Delta$.

For the model eq.(\ref{DP}) on the Bethe lattice with the connectivity $q = \infty$, the self-consistency equations for the local Green's functions are given by \cite{Georges2}
\begin{eqnarray} 
{\cal G}_0(z)^{-1}&=& z -\epsilon_{d} -t_{pd}^2 G_p(z) ,
 \label{SCE3A} \\
G_p(z)^{-1}&=& z - \epsilon_{p} - t_{pd}^2 G_d(z), 
 \label{SCE3B}
\end{eqnarray} 
where $G_p(z)$ is the local Green's function for the $p$-electron and $G_d(z)$ is that for the $d$-electron; 
$\epsilon_{d}\equiv \epsilon_{d0}-\mu=-\mu$ and 
$\epsilon_{p}\equiv \epsilon_{p0}-\mu=\Delta-\mu$. 

In the linearized DMFT, the two-band Hubbard model eq.(\ref{DP}) is mapped onto  the two-site Anderson model eq.(\ref{2SITE}) with $\epsilon_c=0$ and $\epsilon_f=\epsilon_d=-\mu$. In the limit $V\to 0$, the local Green's functions are given by 
$G_d(z) = \frac{Z_d}{z}$ and 
$G_p(z) = \frac{Z_p}{z},$
near the Fermi level with small weights $Z_d \to 0$ and $Z_p \to 0$. Then the self-consistency equations (\ref{SCE3A}) and (\ref{SCE3B}) are reduced to a simple equation 
\begin{equation}
t_{pd}^2 Z_p=V^2. \label{SCE4} 
\end{equation}
To second order in $V$, the quasiparticle weight for the $d$-electron is given by $Z_d=V^2F(U,\mu)$, and that for the $p$-electron is given by \cite{Ono} (see also Appendix~\ref{appendix2})
\begin{eqnarray}
Z_p = A(t_{pd},U,\Delta,\mu)V^2, \label{Z_p}
\end{eqnarray}
with
\begin{eqnarray}
A(t_{pd},U,\Delta,\mu)	= \frac{t_{pd}^2 F(U,\mu)}{
      \left(\Delta-\mu +\frac{t_{pd}^2}{2\mu}
     -\frac{t_{pd}^2}{2(U-\mu)}   \right)^2 }, \label{A}
\end{eqnarray} 
where $F$ is defined in eq.(\ref{F}). 

From eqs.(\ref{SCE4}) and (\ref{Z_p}), we have an equation to determine the MIT point within the linearized DMFT: 
\begin{eqnarray}
t_{pd}^2 A(t_{pd},U,\Delta,\mu)=1. \label{A1}
\end{eqnarray}
As mentioned in \S~\ref{sec:1band}, $V^2$ increases exponentially with iteration number for $t_{pd}^2A>1$ and, then, the single pole approximation for $\Delta(z)$  breaks down resulting in the metallic solution. For $t_{pd}^2A<1$, $V^2$ decreases exponentially to obtain the self-consistent value $V^2 = 0$ corresponding to the insulating solution.

\begin{figure}[ht]
\begin{center}
\leavevmode
\epsfxsize=7.8cm
   \epsffile{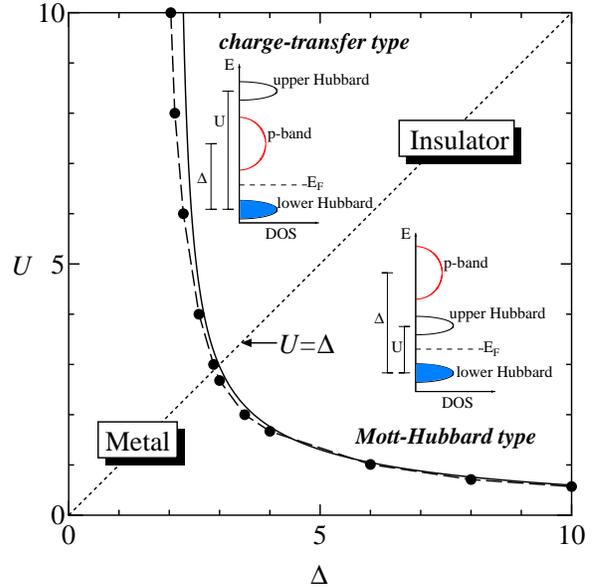}
\caption{
Phase diagram of the two-band Hubbard model at $T=0$ and $n=1$. Solid line is phase boundary separating the metallic and insulating regimes obtained from the linearized DMFT as a function of $\Delta$ and $U$. Closed circles are the critical values of the MIT calculated from the exact diagonalization method \cite{Ono1}. Schematic figures of the density of states are also shown for the Mott-Hubbard type insulator with $U<\Delta$ and the charge-transfer type insulator with $U>\Delta$. 
} 
\label{phase}
\end{center}
\end{figure}

In eq.(\ref{A1}), $A$ includes the chemical potential $\mu$ which has to be determined explicitly to obtain the critical values of the MIT. As shown in the next paragraph, we can use a condition to determine $\mu$, based on the fact that at the MIT point $A$ has a minimum value as a function of $\mu$. This condition gives 
\begin{eqnarray} 
\frac{\partial}{\partial \mu}
    A(t_{pd},U,\Delta,\mu) = 0. 
\label{A2}
\end{eqnarray} 
From the coupled equations (\ref{A1}), (\ref{A2}) with eq.(\ref{A}), we obtain an analytic expression for the phase boundary separating the metallic and insulating regimes as a function of $U$ and $\Delta$ \cite{2band}. Figure. \ref{phase} shows the phase diagram of the two-band Hubbard model at half-filling $n=1$ \cite{mu} on the $\Delta-U$ plane, where $n$ is the electron density per unit cell and given by the sum of $p$- and $d$-electron densities: $n=n_p+n_d$. The analytic results from the linearized DMFT are in good agreement with the available numerical results from the ED method \cite{Ono1} for all values of $\Delta$ and $U$ \cite{Ono}.

When the parameters, $\Delta$ and $U$, are in the metallic regime, the chemical potential $\mu(n)$ is continuous at $n=1$ as a function of $n$. On the other hand, in the insulating regime, $\mu(n)$ has a discontinuity at $n=1$. Correspondingly, there are three cases in the $\mu$ dependence of $A$ as below. 
(1) In the metallic regime, 
$t_{pd}^2A>1$ for all $\mu$ resulting in the metallic solution for all $n$. 
(2) In the insulating regime, 
$t_{pd}^2A<1$ for $\mu_- < \mu < \mu_+$, 
while, $t_{pd}^2A>1$ for $\mu < \mu_-$ or $\mu > \mu_+$. 
Then the system is a Mott insulator for $\mu_- < \mu < \mu_+$, and $\mu$ shows a discontinuity from $\mu_-$ to $\mu_+$ at $n=1$. 
(3) On the phase boundary of the MIT, 
$t_{pd}^2A=1$ for $\mu = \mu(n=1)$, 
while, $t_{pd}^2A>1$ for $\mu \ne \mu(n=1)$. 
Then $A$ has a minimum at $\mu = \mu(n=1)$. 
Therefore the equation (\ref{A2}) is the unique condition to determine the chemical potential on the MIT phase boundary within the linearized DMFT.

\section{Discontinuity in the Chemical Potential}
\label{sec:CP}

\begin{figure}[t]
\begin{center}
\leavevmode
\epsfxsize=7.7cm
   \epsffile{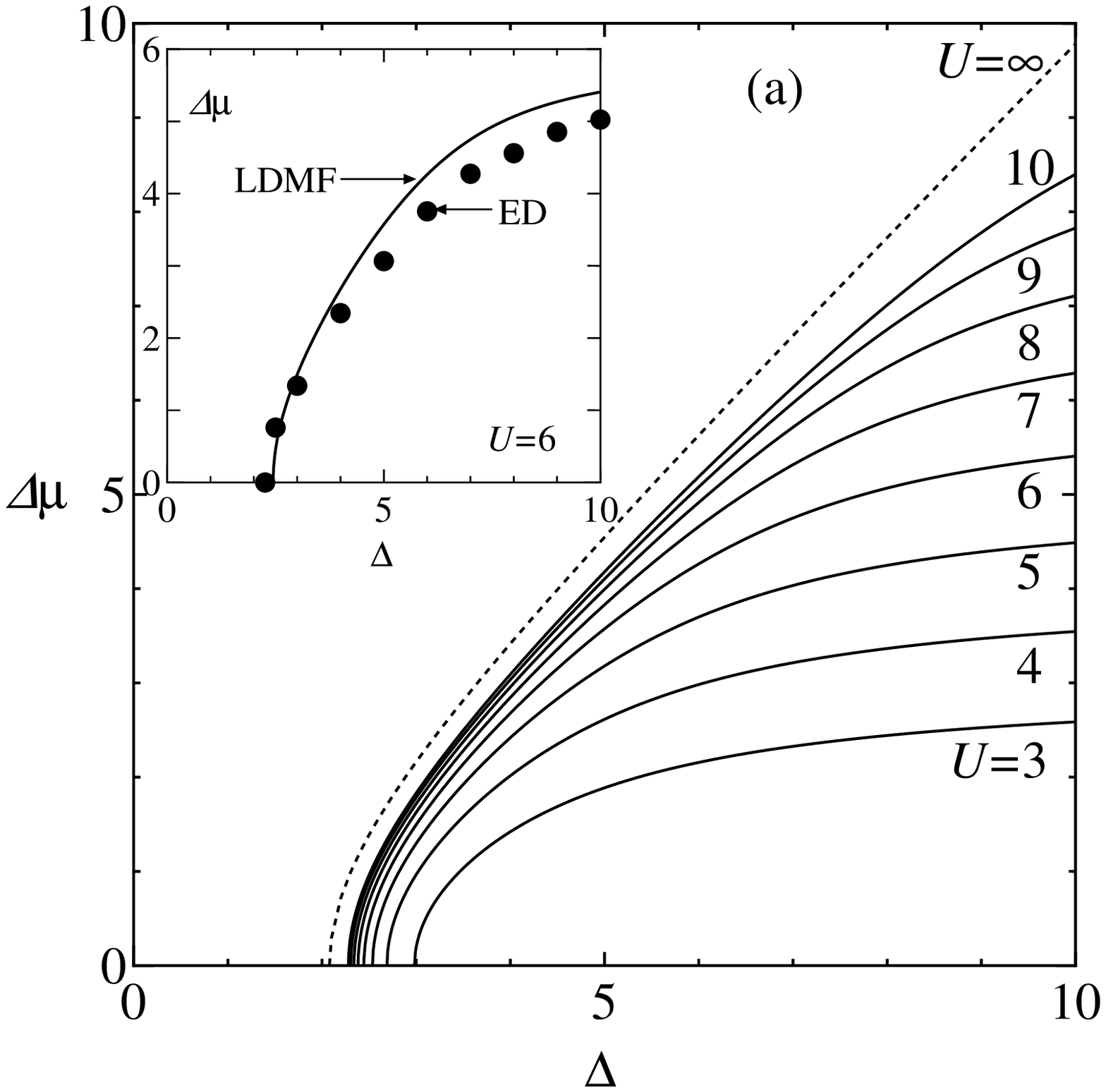}
\vspace{1mm}
\leavevmode
\epsfxsize=7.7cm
   \epsffile{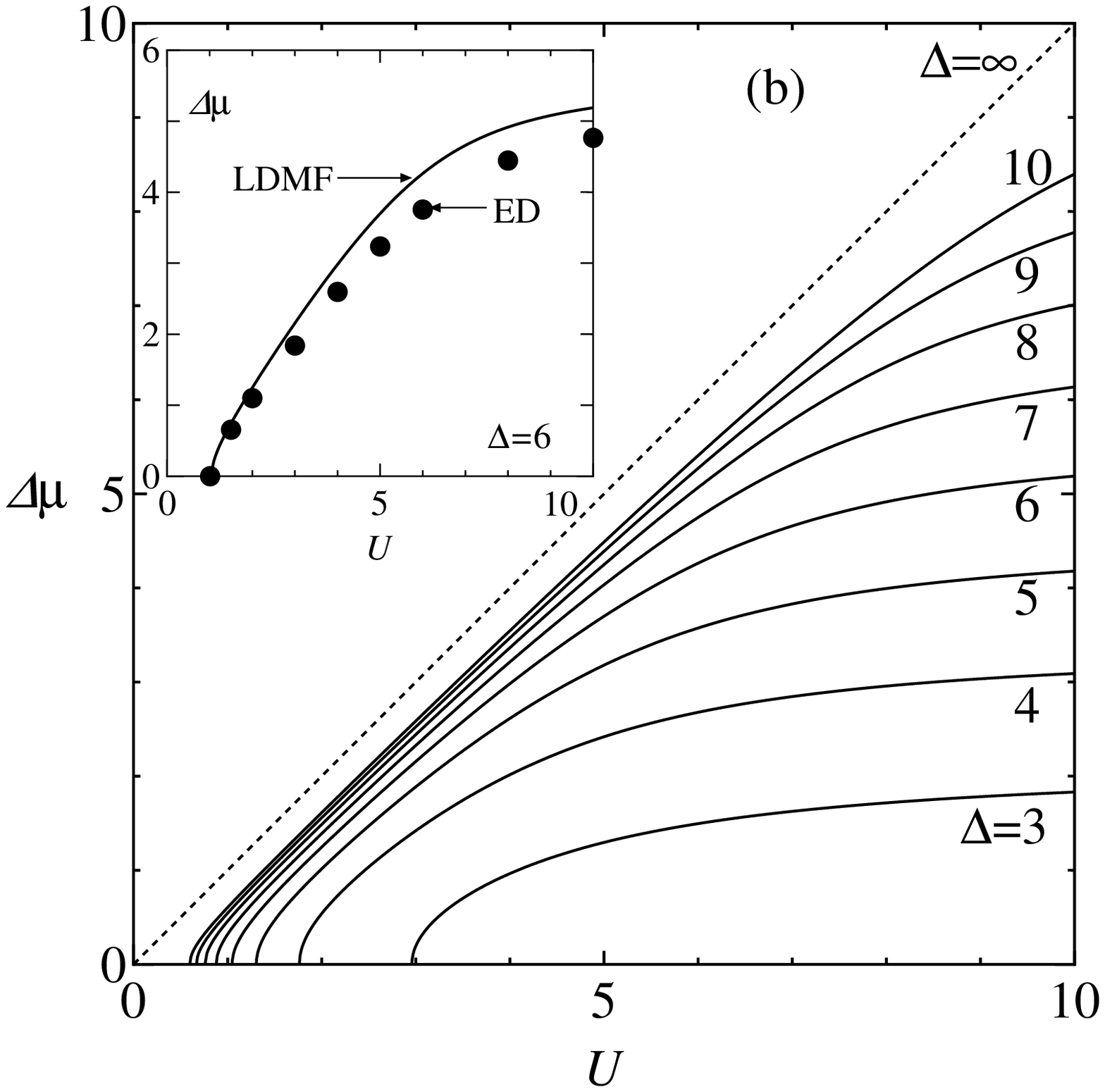}
\caption{
Discontinuity in the chemical potential ${\it \Delta}\mu$ obtained from the linearized DMFT (LDMF) as a function of $\Delta$ for $U=3,4,5,6,7,8,9,10,\infty$ (a) and as a function of $U$ for $\Delta=3,4,5,6,7,8,9,10,\infty$ (b). In the insets, the ED results of ${\it \Delta}\mu$ are also plotted for several $\Delta$ at $U=6$ (a) and for several $U$ at $\Delta=6$ (b).}
\label{Egap}
\end{center}
\end{figure}

When the parameters, $\Delta$ and $U$, are in the insulating regime, the chemical potential has a discontinuity ${\it \Delta}\mu=\mu_+-\mu_-$ at $n=1$ \cite{mu} as mentioned in the previous section, where $\mu_\pm$ is given by the equation 
\begin{eqnarray} 
t_{pd}^2 A(t_{pd},U,\Delta,\mu_\pm)=1. \label{A3}
\end{eqnarray} 
Using eq.(\ref{A}) with eq.(\ref{F}), we can solve eq.(\ref{A3}) to get analytic expressions for $\mu_{\pm}$ as functions of $\Delta$ and $U$, which yield the discontinuity ${\it \Delta}\mu=\mu_+-\mu_-$ within the linearized DMFT. The results of ${\it \Delta}\mu$ are plotted as functions of $\Delta$ for several values of $U$ in Fig.~\ref{Egap}(a) and as functions of $U$ for several values of $\Delta$ in Fig.~\ref{Egap}(b).

In the charge-transfer regime with $U>\Delta$, the MIT occurs at a critical value $\Delta_c(U)$ when $\Delta$ is varied for a fixed $U$ as seen in Fig.~\ref{phase}. In the limit $U\to \infty$, the critical value $\Delta_c$ and the discontinuity in the chemical potential ${\it \Delta}\mu$ are obtained by (see Appendix~\ref{appendix3})
\begin{eqnarray}
 \Delta_c  &= & 2.08t_{pd}, \label{Dc} \\
 {\it \Delta}\mu &=& \sqrt{\Delta^2-\Delta_c^2}, \ \ ({\rm for} \  \Delta>\Delta_c). 
             \label{Dmu}
\end{eqnarray}
The discontinuity ${\it \Delta}\mu$, eq.(\ref{Dmu}), shows a square root dependence of $(\Delta-\Delta_c)$ near $\Delta_c$, while it is given roughly by $\Delta$ for $\Delta\gg\Delta_c$. The similar properties are also observed for finite values of $U$ in the CT regime $U>\Delta$ as seen in Fig.~\ref{Egap}(a).

In the Mott-Hubbard regime with $\Delta>U$, the MIT occurs at a critical value $U_c(\Delta)$ when $U$ is varied for a fixed $\Delta$ as seen in Fig.~\ref{Egap}(b). In the limit $\Delta\to \infty$, the critical value $U_c$ and the discontinuity in the chemical potential ${\it \Delta}\mu$ are obtained by (see Appendix~\ref{appendix3})
\begin{eqnarray}
 U_c  &=&  5.84 \; \frac{t_{pd}^2}{\Delta} \to 0, \\
 {\it \Delta}\mu &=& U.    \label{DU}
\end{eqnarray}
The discontinuity ${\it \Delta}\mu$ is given roughly by $U$ for $U\gg U_c$ even in the case with finite $\Delta$ in the MH regime as seen in Fig.~\ref{Egap}(b). 

\begin{figure}[ht]
\begin{center}
\leavevmode
\epsfxsize=7.4cm
   \epsffile{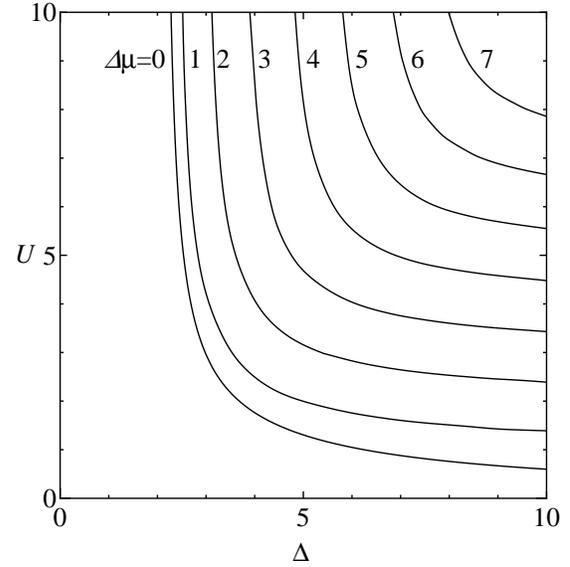}
\caption{
Contour map for ${\it \Delta}\mu$ on the $\Delta-U$ plane calculated from the linearized DMFT.
}
\label{Contour}
\end{center}
\end{figure}

%
%
%
The above mentioned features of the discontinuity in the chemical potential for both the CT and MH regimes can be clearly seen in the contour map for ${\it \Delta}\mu$ on the $\Delta-U$ plane shown in Fig.~\ref{Contour}. Note that the line with ${\it \Delta}\mu=0$ is equivalent to the phase boundary of the MIT shown in Fig.~\ref{phase}.

%
%
We have also calculated ${\it \Delta}\mu$ numerically by using the ED method. We solved the full DMFT equation numerically and obtained the electron density $n$ as a function of $\mu$. When $\mu$ approaches $\mu_+$ ($\mu_-$) from above (below), $n$ approaches unity from above (below) and $n=1$ for $\mu_-<\mu<\mu_+$. We also confirmed that, at $\mu=\mu_{\pm}$, the groundstate changes from singlet ($\mu<\mu_-$ or $\mu>\mu_+$) to doublet ($\mu_-<\mu<\mu_+$). The actual calculations were done for finite cluster sizes $n_s$ up to $n_s=10$. The size dependence of ${\it \Delta}\mu$ was very small for $n_s \geq 6$. The results of ${\it \Delta}\mu$ obtained from the ED method with the cluster size $n_s=8$ are also plotted in the insets in Figs.~\ref{Egap}(a) and (b). 

As seen in the insets in Figs.~\ref{Egap}(a) and (b), the analytic results from the linearized DMFT are in good agreement with the numerical results from the ED method. We note that the value of ${\it \Delta}\mu$ from the linearized DMFT seems to be larger than that from the ED for the strong-coupling case. This will be discussed in \S 5.

\section{Critical Behavior near the Mott transition}
\label{sec:critical}

Finally, we discuss the critical behaviour near the Mott MIT at half-filling in the two-band Hubbard model. In this case, we need the result of the quasiparticle weight for the $d$-electron up to fourth order in $V$ in the general case including the non-symmetric case with $\mu\ne\frac{U}{2}$ \cite{Ono3} (see also Appendix~\ref{appendix1}) 
\begin{eqnarray} 
Z_d=F(U,\mu)V^2-G(U,\mu)V^4, \label{Z_d2}
\end{eqnarray} 
where $F$ is given in eq.(\ref{F}) and 
\begin{eqnarray} 
G(U,\mu)&=&\frac{29}{2\mu^4}+\frac{24}{\mu^3(U-\mu)}+\frac{22}{\mu^2(U-\mu)^2}
              \nonumber \\
        & &    +\frac{24}{\mu(U-\mu)^3}+\frac{29}{2(U-\mu)^4}. 
\end{eqnarray} 
The quasiparticle weight for the $p$-electron $Z_p$ is also calculated up to fourth order in $V$ (see Appendix~\ref{appendix2}) 
\begin{eqnarray}
	Z_p   = A(t_{pd},U,\Delta,\mu)V^2 - B(t_{pd},U,\Delta,\mu)V^4  \; , 
	\label{Z_p2}
\end{eqnarray}
where $A$ is given in eq.(\ref{A}) and $B$ is given in the Appendix~\ref{appendix2}. 
Substituting $V^2$ from the self-consistency equation (\ref{SCE4}) into eqs.(\ref{Z_d2}) and (\ref{Z_p2}), we obtain 
\begin{eqnarray}
  Z_d &=& t_{pd}^2F(U,\mu)Z_p ,   \label{Z_d3} \\
  Z_p &=& \frac{t_{pd}^2A(t_{pd},U,\Delta,\mu)-1}
	           {t_{pd}^4B(t_{pd},U,\Delta,\mu)} , \label{Z_p3} 
\end{eqnarray}
in the metallic regime close to the MIT phase boundary. 

At the MIT point with the critical values $U_c$, $\Delta_c$ and $\mu_c$, eqs.(\ref{Z_d3}) and (\ref{Z_p3}) yield $Z_d=Z_p=0$ from eq.(\ref{A1}).
When $\Delta$ or $U$ decreases from the MIT point, 
the quasiparticle weight for the $d$-electron $Z_d$ eq.(\ref{Z_d3}) and that for the $p$-electron $Z_p$ eq.(\ref{Z_p3}) increase 
as
\begin{eqnarray}
  Z_\nu &=& C^\nu_{\Delta} \left(1-\frac{\Delta}{\Delta_c}\right), \ \ \
          (\nu=d \ {\rm or} \ p)         \label{ZdD} \\
  Z_\nu &=& C^\nu_{U} \left(1-\frac{U}{U_c}\right),  \ \ \ 
          (\nu=d \ {\rm or} \ p)        \label{ZdU} 
\end{eqnarray}
near the MIT point at half-filling, respectively, where the coefficients are given by 
\begin{eqnarray}
\hspace*{-8mm}C_{\Delta}^d &=& 
      -\frac{\Delta_c F(U_c,\mu_c)}{B(t_{pd},U_c,\Delta_c,\mu_c)}
       \frac{\partial A(t_{pd},U_c,\Delta,\mu_c)}{\partial \Delta}
       \bigg|_{\Delta=\Delta_c}\hspace{-1mm} ,     \label{alphaD} \\
\hspace*{-8mm}C_{U}^d &=& 
      -\frac{U_c F(U_c,\mu_c)}{B(t_{pd},U_c,\Delta_c,\mu_c)}
       \frac{\partial A(t_{pd},U,\Delta_c,\mu_c)}{\partial U}
       \bigg|_{U=U_c} , \hspace{-5mm}  \label{alphaU}  \\
\hspace*{-8mm}\frac{C_{\Delta}^p}{C_{\Delta}^d}&=& \frac{C_{U}^p}{C_{U}^d} =
      \frac{t_{pd}^2}{ \left(\Delta_c-\mu_c +\frac{t_{pd}^2}{2\mu_c}
     -\frac{t_{pd}^2}{2(U_c-\mu_c)} \right)^2}  . \label{ratio}
\end{eqnarray}

\begin{figure}[ht]
\begin{center}
\leavevmode
\epsfxsize=7cm
   \epsffile{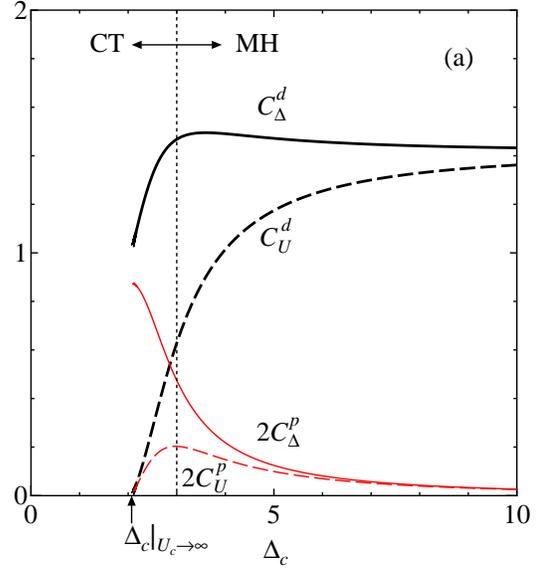}
\vspace{5mm}
\leavevmode
\epsfxsize=7cm
   \epsffile{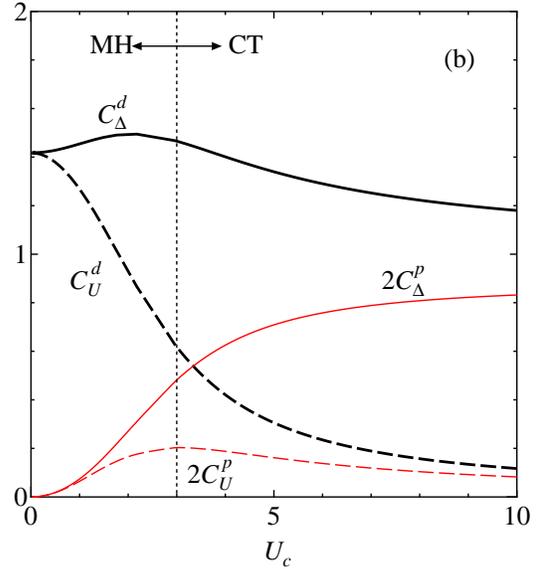}
\caption{
The coefficients $C_{\Delta}^d$ (thick solid line), $C_{U}^d$ (thick dashed line), $C_{\Delta}^p$ (thin solid line) and $C_{U}^p$ (thin dashed line) of the quasiparticle weights for the $d$ and $p$-electrons 
$
Z_{d(p)}= C_{\Delta}^{d(p)}\left(1-\frac{\Delta}{\Delta_c}\right) 
$
and 
$
Z_{d(p)} = C_{U}^{d(p)} \left(1-\frac{U}{U_c}\right), 
$
respectively, as functions of $\Delta_c$ (a) and $U_c$ (b). 
}
\label{alpha}
\end{center}
\end{figure}

Figures. \ref{alpha}(a) and (b) show the coefficients $C_\Delta^d$, $C_{U}^d$, $C_\Delta^p$ and $C_{U}^p$ given in eqs.(\ref{alphaD})-(\ref{ratio}). 
In the CT regime with $U>\Delta$, the critical values are $\Delta_c\siml 3t_{pd}$ and $U_c\simg 3t_{pd}$, while, in the MH regime with $U<\Delta$, they are $\Delta_c\simg 3t_{pd}$ and $U_c\siml 3t_{pd}$ (see Fig.~\ref{phase}).

In the CT regime, the MIT occurs at $\Delta=\Delta_c$ for a fixed $U\simg 3t_{pd}$. As $\Delta$ decreases below $\Delta_c$ for a fixed $U$, the quasiparticle weight increases as given in eq.(\ref{ZdD}). With increasing $U_c$ (decreasing $\Delta_c$), the coefficient $C_\Delta^d$ decreases due to the increasing correlation effect, while $C_\Delta^p$ increases because of the rapid increase in the $p$-component for the quasiparticle weight $Z_p/Z_d=C_\Delta^p/C_\Delta^d$ near the MIT. In the limit $U_c\to\infty$ ($\Delta_c\to 2.08t_{pd}$), we find $C_{\Delta}^d = 1.01$, $C_{U}^d= 0$ and $C_\Delta^p/C_\Delta^d=C_U^p/C_U^d=0.432$ (see Appendix~\ref{appendix3}).

In the MH regime, the MIT occurs at $U=U_c$ for a fixed $\Delta\siml 3t_{pd}$. As $U$ decreases below $U_c$ for a fixed $\Delta$, the quasiparticle weight increases as given in eq.(\ref{ZdU}). With increasing $\Delta_c$ (decreasing $U_c$), the coefficient $C_U^d$ increases due to the decreasing correlation effect, while $C_U^p$ decreases because of the rapid decrease in the $p$-component for the quasiparticle weight $Z_p/Z_d=C_U^p/C_U^d$ near the MIT. In the limit $\Delta_c\to\infty$ ($U_c\to 0$), we find $C_{\Delta}^d = 1.42$, $C_{U}^d= 1.42$ and $C_\Delta^p/C_\Delta^d=C_U^p/C_U^d=0$ (see Appendix~\ref{appendix3}). 
We note that, even in the limit $\Delta_c\to\infty$, the effect of the $p$-band is still relevant ($C_{\Delta}^d$ is finite), because, the hopping integrals between $p$-$p$ and $d$-$d$ orbitals were not considered in the present model eq.(\ref{DP}) and, then, the electron has to transfer between $d$ and $p$ orbitals through the hopping integral $t_{pd}$ \cite{2band}.

\begin{figure}[t]
\begin{center}
\leavevmode
\epsfxsize=7.5cm
   \epsffile{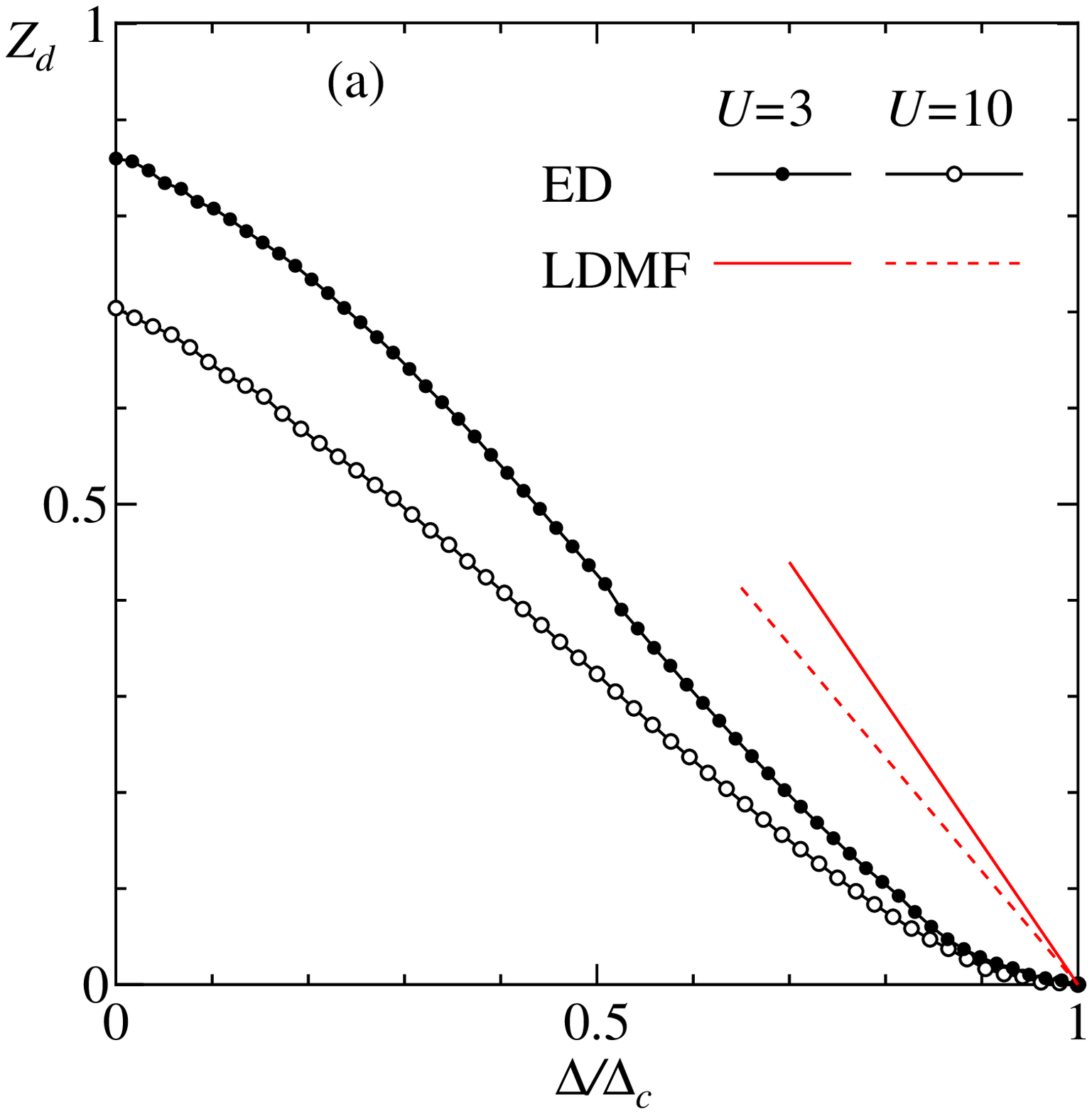}
\vspace{5mm}
\leavevmode
\epsfxsize=7.5cm
   \epsffile{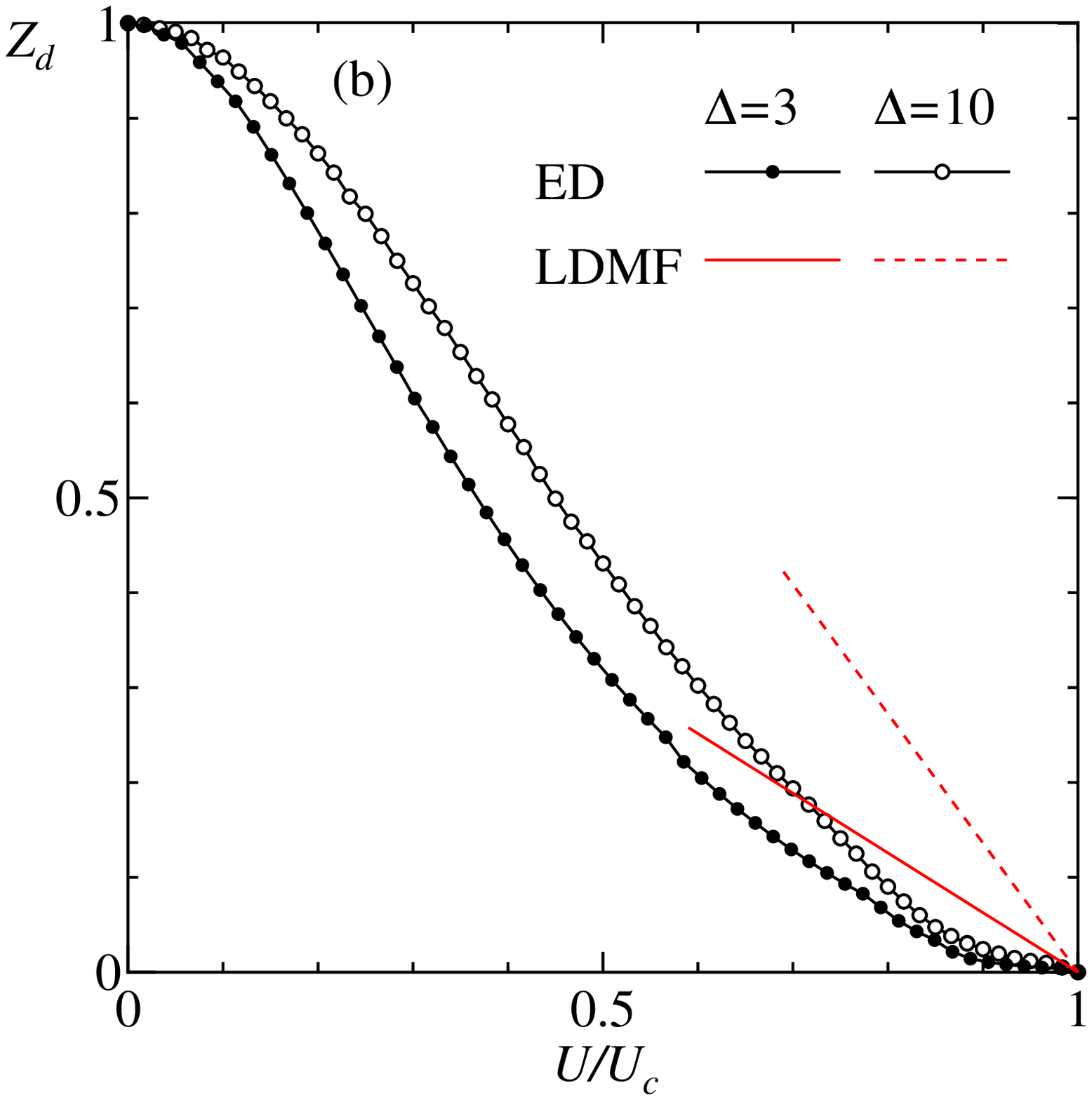}
\caption{
Quasiparticle weight for the $d$-electron $Z_d$ calculated from the ED method together with that from the linearized DMFT as functions of $\Delta/\Delta_c$ for $U=3$ ($\Delta_c^{ED}=2.95$) and $U=10$ ($\Delta_c^{ED}=2.08$) (a), and as functions of $U/U_c$ for $\Delta=3$ ($U_c^{ED}=2.65$) and $\Delta=10$ ($U_c^{ED}=0.60$) (b). 
}
\label{Zd}
\end{center}
\end{figure}

We have also calculated the quasiparticle weight for the $d$-electron 
$ 
Z_d=(1-\frac{d\Sigma (z)}{d z}|_{z=0} )^{-1},
$ 
with the local self-energy
$ 
\Sigma (z)={\cal G}_0(z)^{-1} - G_d(z)^{-1}, 
$ 
numerically by using the ED method. The coefficients $C_{\Delta}^d$ and $C_{U}^d$ obtained from the linearized DMFT seem to be about five times larger than those from the ED method for all parameter regimes as seen in Figs.~\ref{Zd}(a) and (b). The similar discrepancy has been found in the single-band Hubbard model mentioned in \S 2.1 \cite{Bulla}. The trends, however, within the both methods are consistent with each other as follows:  (1) In the CT regime, $C_{\Delta}^d$ decreases with increasing $U$ (see Fig.\ref{Zd}(a)). (2) In the MH regime, $C_{U}^d$ increases with increasing $\Delta$ (see Fig.\ref{Zd}(b)). (3) In the intermediate regime $U_c \sim \Delta_c \sim 3t_{pd}$, $C_{\Delta}^d$ is about twice larger than $C_{U}^d$ (see Figs.\ref{Zd}(a) and (b)). (4) $C_{\Delta}^d$ in the CT regime with $U=10 t_{pd}$ is almost the same as $C_{U}^d$ in the MH regime with $\Delta=10 t_{pd}$ (see Figs.\ref{Zd}(a) and (b)). 
Further improvements in both the analytical and numerical methods are under way to obtain a more conclusive description for the critical behaviour near the Mott MIT, which will be reported in a future publication.

\section{Summary and Discussion}
\label{sec:disc}

Within the linearized DMFT, we have studied the Mott metal-insulator transition in the two-band Hubbard model characterized by the two parameters: the $d$-site Coulomb interaction $U$ and the $d$-$p$ charge-transfer energy $\Delta$, and obtained a unified description of the MIT analytically over the whole parameter regime including the Mott-Hubbard regime, the charge-transfer regime and the intermediate regime as follows: 

(1) The Mott-Hubbard regime ($U<\Delta$) : 
The MIT occurs at a critical value $U_c$ when $U$ is varied for a fixed $\Delta$ at half-filling. The critical value $U_c$ monotonically decreases with increasing $\Delta$. When $U$ increases above $U_c$ for a fixed $\Delta$, the discontinuity in the chemical potential ${\it \Delta}\mu$ increases: ${\it \Delta}\mu \propto \sqrt{U-U_c}$ near $U_c$ and ${\it \Delta}\mu \sim U$ for $U\gg U_c$. When $U$ decreases below $U_c$ for a fixed $\Delta$, the quasiparticle weight for the $d$-electron $Z_d$ and that for $p$-electron $Z_p$ increase as $Z_d=C^d_U(1-\frac{U}{U_c})$ and $Z_p=C^p_U(1-\frac{U}{U_c})$ near $U_c$, respectively, where the coefficient $C^d_U$ ($C^p_U$) monotonically increases (decreases) with increasing $\Delta$. 

(2) The charge-transfer regime ($U>\Delta$) : 
The MIT occurs at $\Delta_c$ when $\Delta$ is varied for a fixed $U$, where $\Delta_c$ monotonically decreases with increasing $U$. When $\Delta$ increases above $\Delta_c$ for a fixed $U$, ${\it \Delta}\mu$ increases as ${\it \Delta}\mu \propto \sqrt{\Delta-\Delta_c}$ near $\Delta_c$ and ${\it \Delta}\mu \sim \Delta$ for $\Delta\gg \Delta_c$. When $\Delta$ decreases below $\Delta_c$ for a fixed $U$,  $Z_d$ and $Z_p$ increase as $Z_d=C^d_\Delta(1-\frac{\Delta}{\Delta_c})$ and $Z_p=C^p_\Delta(1-\frac{\Delta}{\Delta_c})$ near $\Delta_c$, respectively, where the coefficient $C^d_\Delta$ ($C^p_\Delta$) monotonically decreases (increases) with increasing $U$. 

(3) The intermediate regime ($U\sim\Delta$) : 
The critical behaviour near the transition as well as the phase boundary smoothly connects the MH type and the CT type. As $U$ increases ($\Delta$ decreases), the $p$-component for the quasiparticle weight $Z_p/Z_d$ near the MIT rapidly increases, which shows the change in the character of the quasiparticle from $d$-band like in the MH regime to $p$-band like in the CT regime. 

We have estimated the reliability of the linearized DMFT by comparing the present analytical results with the available numerical results from the exact diagonalization method. The analytical result for ${\it \Delta}\mu$ near the MIT as well as that for the critical value of the MIT \cite{Ono} is in very good agreement with the numerical result within the ED method. In the strong coupling regime, however, the value of ${\it \Delta}\mu$ from the linearized DMFT seems to be larger than that from the ED method. This may be explained by the fact that here the chemical potential is very close to the edge of the Hubbard bands and/or the $p$-band, and that the effect of the bandwidth, which is neglected in the linearized DMFT, becomes important \cite{Ono3}. 

The analytic results for the coefficients $C^d_\Delta$ and $C^d_U$ from the linearized DMFT seem to be considerably larger than the available ED results, although a precise value within the purely numerical methods has not been obtained yet. When $U$ and/or $\Delta$ decreases form the critical point of the MIT, not only the weight but also the width of the quasiparticle peak is found to increase near the MIT \cite{Bulla1}. The width effect, which is not taken into account in the linearized DMFT, may cause this discrepancy. Further improvements in both the analytical and the numerical methods are now under way. 

Effects of the hopping integrals between $p$-$p$ and $d$-$d$ orbitals, which are not considered in the present study but are not negligible in actual compounds, make important contribution to the critical behaviour near the MIT, as found to be significant to determine the phase boundary of the MIT \cite{Ono}. Furthermore, the effects of the doping, which is not discussed here but has been discussed in the single-band Hubbard model \cite{Ono3}, make apparent differences between the hole doping and electron doping cases in the strong coupling CT regime, where the chemical potential is just below the $p$-band for the electron doping while it is just above the lower-Hubbard band for the hole doping. Such effects and the improvements of the methods will be reported in a future publication.

\vspace{5mm}

{\flushleft{\bf Acknowledgments}}

\vspace{2mm}

One of the authors (Y\=O) would like to thank  A. C. Hewson,  R. Bulla and  M. Potthoff for valuable discussions. This work was partially supported by the Grant-in-Aid for Scientific Research from the Ministry of Education, Science, Sports and Culture, and also by CREST (Core Research for Evolutional Science and Technology) of Japan Science and Technology Corporation (JST).

\appendix
\section{Two-site Anderson Model}
\label{appendix1}

Here we discuss the two-site Anderson model eq.(\ref{2SITE}) \cite{Hewson}. We assume that the conduction level is between the atomic $f$-level and the upper-Hubbard level 
$
\epsilon_f<\epsilon_{c} < \epsilon_f+U, 
$
and we set $\epsilon_{c}= 0$ and $\epsilon_{f}= -\mu$. Then we define 
\begin{eqnarray} 
 X &\equiv& \epsilon_{c}-\epsilon_f  = \mu >0, \nn\\
 Y &\equiv& \epsilon_f+U -\epsilon_{c} = U-\mu >0. \nn
\end{eqnarray} 
Henceforth we consider the case with the small hybridization strength $|V| \ll X,Y$, and calculate all the quantities up to fourth order in $V$ \cite{Ono3}.

The one-electron eigenenergies and the corresponding eigenstates are
\begin{eqnarray} 
E_{+} &=&  \epsilon_{c} +\frac{V^2}{X}-\frac{V^4}{X^3}, \nn\\ 
E_{-} &=&  \epsilon_f -\frac{V^2}{X}+\frac{V^4}{X^3}, \nn\\
|E_{+}\rangle &=&  \alpha\left\{\frac{V}{X}\left(1-\frac{V^2}{X^2} \right) 
                 f_\sigma^+ + c_\sigma^+  \right\} |0\rangle,    \nn\\
|E_{-}\rangle &=&  \alpha\left\{f_\sigma^+ - \frac{V}{X} 
          \left(1-\frac{V^2}{X^2} \right)  c_\sigma^+  \right\} |0\rangle,\nn
\end{eqnarray} 
where
$\alpha^2 = 1-\frac{V^2}{X^2} +3\frac{V^4}{X^4}$. 

Similarly, the three-electron (one-hole) eigenenergies and the corresponding eigenstates are
\begin{eqnarray} 
\bar{E}_{+} &=&  \epsilon_{c} +2\epsilon_f+U 
               +\frac{V^2}{Y}-\frac{V^4}{Y^3}, \nn\\
\bar{E}_{-} &=&  2\epsilon_{c}+\epsilon_f 
               -\frac{V^2}{Y}+\frac{V^4}{Y^3},\nn\\
|\bar{E}_{+}\rangle &=&  \bar{\alpha}\left\{-\frac{V}{Y}\left(1-\frac{V^2}{Y^2}                  \right)  f_\sigma + c_\sigma  \right\} f_\uparrow^+ 
                 f_\downarrow^+ c_\uparrow^+ c_\downarrow^+|0\rangle,   \nn\\
|\bar{E}_{-}\rangle &=&  \bar{\alpha}\left\{f_\sigma + \frac{V}{Y} 
          \left(1-\frac{V^2}{Y^2} \right)  c_\sigma  \right\} f_\uparrow^+ 
          f_\downarrow^+ c_\uparrow^+ c_\downarrow^+|0\rangle,\nn
\end{eqnarray} 
where 
$\bar{\alpha}^2 = 1-\frac{V^2}{Y^2} +3\frac{V^4}{Y^4}$. 

The ground state for two electron is the singlet state with the following eigenenergy and the eigenstate
\begin{eqnarray} 
\hspace{-0.5cm} E_{0} &=&    \epsilon_{c} +\epsilon_f  -2V^2 \left(\frac{1}{X}+\frac{1}{Y} \right)  \left(1-\frac{2V^2}{X^2}-\frac{2V^2}{Y^2} \right), \nn\\
\hspace*{-0.5cm} |E_{0}\rangle 
     &=& \alpha_0\left\{\frac{1}{\sqrt{2}} (c_\uparrow^+ f_\downarrow^+  
     - c_\downarrow^+ f_\uparrow^+ )|0\rangle \right.\nn\\ \hspace*{-1cm} 
     & & \hspace{0.3cm} -\frac{\sqrt{2}V}{X} \left(1-\frac{2V^2}{X^2} 
     -\frac{2V^2}{X Y} \right)  c_\uparrow^+ c_\downarrow^+ |0\rangle   
     \nonumber \\ \hspace*{-1cm}     
     & & \hspace{0.3cm}  \left.-\frac{\sqrt{2}V}{Y}  \left(1-\frac{2V^2}{Y^2}
     -\frac{2V^2}{X Y} \right)  f_\uparrow^+ f_\downarrow^+ |0\rangle
     \right\} , \label{E0}\nn
\end{eqnarray}
where 
$
\alpha_0^2 = 1-2V^2(\frac{1}{X^2}+\frac{1}{Y^2} ) 
           {}+4V^4(\frac{3}{X^4}+\frac{2}{X^3 Y}+\frac{2}{X^2 Y^2} 
             +\frac{2}{X Y^3}+\frac{3}{Y^4}).
$

When a $f,\uparrow$ electron is removed from the ground state $|E_{0}\rangle$, there are two possible final states: $|E_{+}\rangle$ and $|E_{-}\rangle$. Correspondingly, there are two possible single-hole excitations with excitation energies, 
\begin{eqnarray} 
E_+-E_0 &\equiv& -\epsilon_f+c_{1\epsilon 2}V^2 - c_{1\epsilon 4}V^4    \equiv -\epsilon_1, \label{E1} \\
E_--E_0 &\equiv& c_{2\epsilon 2}V^2 - c_{2\epsilon 4}V^4    \equiv -\epsilon_2, \label{E2} 
\end{eqnarray}
where
\begin{eqnarray}
c_{1\epsilon 2}        &=&\frac{3}{X}+\frac{2}{Y},  \nn\\
c_{1\epsilon 4}        &=&\frac{5}{X^3}+\frac{4}{X^2 Y}+\frac{4}{X Y^2}+\frac{4}{Y^3} ,\nn\\
c_{2\epsilon 2} &=& \frac{1}{X}+\frac{2}{Y},   \nn\\
c_{2\epsilon 4} &=& \frac{3}{X^3}+\frac{4}{X^2 Y}+\frac{4}{X Y^2}+\frac{4}{Y^3}.\nn
\end{eqnarray}
And the transition probabilities are calculated as,
\begin{eqnarray} 
\hspace*{-7mm}
|\langle E_+|f_\uparrow|E_0\rangle|^2 &\equiv& \frac{1}{2}-c_{1w2}V^2 + c_{1w4}V^4 \equiv w_1,  \\
\hspace*{-7mm}
|\langle E_-|f_\uparrow|E_0\rangle|^2 &\equiv& c_{2w2}V^2 - c_{2w4}V^4 \equiv w_2, 
\end{eqnarray}
where
\begin{eqnarray}
c_{1w2}      &=& \frac{1}{2}\left(\frac{3}{X^2}+\frac{4}{X Y}+\frac{2}{Y^2}  
             \right) ,  \nn \\
c_{1w4}      &=& \frac{1}{2} \left(\frac{17}{X^4}+\frac{24}{X^3 Y} +\frac{22}
             {X^2 Y^2}+\frac{24}{X Y^3}+\frac{12}{Y^4}\right) ,  \nn\\
c_{2w2}      &=& \frac{1}{2}\left(\frac{1}{X^2}+\frac{4}{X Y}+\frac{4}{Y^2}
             \right) ,\nn\\
c_{2w2}      &=& \frac{1}{2} \left(\frac{5}{X^4}+\frac{16}{X^3 Y}+
             \frac{22}{X^2 Y^2}+\frac{32}{X Y^3}+\frac{24}{Y^4}\right) . \nn
\end{eqnarray}

Similarly, there are two possible single-particle excitations with excitation energies, 
\begin{eqnarray} 
\bar{E}_--E_0   &\equiv& c_{3\epsilon 2}V^2 - c_{3\epsilon 4}V^4 
                \equiv \epsilon_3, \label{E3}\\
\bar{E}_+-E_0   &\equiv& \epsilon_f+U+c_{4\epsilon 2}V^2 - c_{4\epsilon 4}V^4 
                \equiv \epsilon_4, \label{E4} 
\end{eqnarray}
where
\begin{eqnarray}
c_{3\epsilon 2} &=& \frac{2}{X}+\frac{1}{Y}, \nn \\
c_{3\epsilon 4} &=& \frac{4}{X^3}+\frac{4}{X^2 Y}
                    +\frac{4}{X Y^2}+\frac{3}{Y^3},\nn\\
c_{4\epsilon 2} &=& \frac{2}{X}+\frac{3}{Y} ,  \nn\\
c_{4\epsilon 4} &=& \frac{4}{X^3}+\frac{4}{X^2 Y}+\frac{4}{X Y^2} 
                    +\frac{5}{Y^3}. \nn
\end{eqnarray}
And the transition probabilities are calculated as,
\begin{eqnarray} 
\hspace*{-7mm}|\langle \bar{E}_-|f_\uparrow|E_0\rangle|^2  
               &\equiv& c_{3w2}V^2 - c_{3w4}V^4 \equiv w_3, \\
\hspace*{-7mm}|\langle \bar{E}_+|f_\uparrow|E_0\rangle|^2 
               &\equiv& \frac{1}{2}-c_{4w2}V^2 + c_{4w4}V^4 \equiv w_4,  
               \label{Y4}
\end{eqnarray}
where
\begin{eqnarray}
c_{3w2}      &=& \frac{1}{2}\left(\frac{4}{X^2}+\frac{4}{X Y}+\frac{1}{Y^2} 
             \right) , \nn\\
c_{3w4}      &=& \frac{1}{2} \left(\frac{24}{X^4}+\frac{32}{X^3 Y}+\frac{22}{X^2 Y^2}+\frac{16}{X Y^3}+\frac{5}{Y^4}\right) ,  \nn\\
c_{4w2}      &=& \frac{1}{2}\left(\frac{2}{X^2}+\frac{4}{X Y}+\frac{3}{Y^2}\right), \nn\\
c_{4w4}      &=& \frac{1}{2} \left(\frac{12}{X^4}+\frac{24}{X^3 Y}+\frac{22}{X^2 Y^2}+\frac{24}{X Y^3}+\frac{17}{Y^4}\right) . \nn
\end{eqnarray}

From eqs.(\ref{E1}-\ref{Y4}), we obtain the $f$-electron Green's function which has four poles, 
$
G_\sigma(z) = \sum_{i=1}^4 \frac{w_i}{z-\epsilon_i}. 
$
In the limit $V\to 0$, high-energy poles at $\epsilon_1 \approx \epsilon_f$ and $\epsilon_4 \approx \epsilon_f+U$ have large residues $w_1 \approx w_2 \approx \frac12$, while low-energy poles merge together at $\epsilon_2 \approx \epsilon_3 \approx 0$ with small total weight $Z \equiv w_2+w_3$: 
\begin{eqnarray} 
  Z &=& (c_{2w2} + c_{3w2}) V^2 - (c_{2w4} + c_{3w4}) V^4\nn\\
  &=& F V^2 - G V^4  \label{Z}
\end{eqnarray}
to fourth order in $V$, where
\begin{eqnarray}
F &=& \frac{5}{2X^2}+\frac{4}{X Y}+\frac{5}{2Y^2} ,\nn\\
G &=& \frac{29}{2X^4}+\frac{24}{X^3 Y} 
            +\frac{22}{X^2 Y^2}+\frac{24}{X Y^3}+\frac{29}{2Y^4} .\nn
\end{eqnarray}

\section{Calculation of $Z_p$}
\label{appendix2}
In the limit $V\to 0$, the local $d$-Green's function $G_d(z)$ is approximately given by the three-pole function (see Appendix\ref{appendix1})
\begin{eqnarray}
G_d(z)\cong\frac{w_1}{z-\epsilon_1}+\frac{Z_d}{z}+\frac{w_4}{z-\epsilon_4}, 
\label{G_d}
\end{eqnarray}
where $Z_d\equiv Z$ given in eq.(\ref{Z}). 
Substituting eq.(\ref{G_d}) into eq.(\ref{SCE3B}), $G_p(z)$ is obtained as a four-pole function. In the limit $V\to 0$, $G_p$ is written by 
$
G_p(z)\cong\frac{Z_p}{z-{\it \Delta}\epsilon}
$
near the Fermi level with the small weight $Z_p\to 0$ and the small energy ${\it \Delta}\epsilon\to 0$. The energy ${\it \Delta}\epsilon$ is calculated from $G_p^{-1}({\it \Delta}\epsilon)=0$ with eqs.(\ref{SCE3B}) and (\ref{G_d}):
\begin{eqnarray}
\hspace{-5mm}{\it \Delta}\epsilon&=&-t_{pd}^2{c_{2w2} + c_{3w2}   \over E_{p}}V^2 
+ t_{pd}^2\bigg\{{c_{2w4} + c_{3w4}\over E_{p}} \nn\\
\hspace{-5mm} &+& {B_1 \over E_{p}^2} (c_{2w2} + c_{3w2})
+{B_2\over E_{p}^3} (c_{2w2} +  c_{3w2} )^2     \bigg\} V^4 , \label{dele}
\end{eqnarray}
to fourth order in $V$, where
\begin{eqnarray}
	E_p &=& \Delta-\mu-t_{pd}^2\left(-\frac{1}{2\mu}+\frac{1}{2(U-\mu)}\right)  ,\nn\\
    B_1 &=& -\frac{c_{1\epsilon 2}}{2\mu^2}-\frac{c_{1w2}}{\mu}+\frac{c_{4\epsilon 2}}{2(U-\mu)^2}+\frac{c_{4w2}}{U-\mu},\nn\\
	B_2 &=& 1+t_{pd}^2\left(\frac{1}{2\mu^2}+\frac{1}{2(U-\mu)^2}\right) .\nn
\end{eqnarray}
The residue $Z_p$ is calculated from 
$Z_p^{-1} = \frac{d}{dz} G_p^{-1}(z)|_{z={\it \Delta}\epsilon}$ 
with eqs.(\ref{SCE3B}), (\ref{G_d}) and (\ref{dele}):
\begin{eqnarray}
	Z_p   &=& A(t_{pd},U,\Delta,\mu)V^2 - B(t_{pd},U,\Delta,\mu)V^4  \; , 
\end{eqnarray}
to fourth order in $V$, where the coefficients are
\begin{eqnarray}
A(t_{pd},U,\Delta,\mu)	&=& t_{pd}^2 \frac{c_{2w2} + c_{3w2}}{
      E_p^2 } ,\\
	B(t_{pd},U,\Delta,\mu)  &=& t_{pd}^2\frac{c_{2w4}+c_{3w4}}{E_p^2} 
+t_{pd}^4  \frac{2B_1 (c_{2w2}+c_{3w2})}{E_p^3}\nn\\
&&\hspace*{0cm}
+t_{pd}^4 \frac{3B_2(c_{2w2}+c_{3w2})^2}{E_p^4}
.  \label{B}
\end{eqnarray}

\section{Two Limiting Cases: $U\to\infty$ and $\Delta\to\infty$}
\label{appendix3}

In the limit $U\rightarrow \infty$, eqs. (\ref{A1}) and (\ref{A2}) are reduced to
\begin{eqnarray}
	5t_{pd}^4 - 2\left( \Delta-\mu+\frac{t_{pd}^2}{2\mu}\right)^2 \mu^2 
	&=& 0 ,  \label{Uinfty}\\
	\Delta-\mu+\frac{t_{pd}^2}{2\mu}-\left(1+\frac{t_{pd}^2}{4\mu^2}\right)\mu 
    &=& 0 , \label{Uinfty1}
\end{eqnarray}
respectively. By solving the coupled equations (\ref{Uinfty}) and (\ref{Uinfty1}), we obtain the critical values of the MIT 
\begin{eqnarray}
	\mu_{c} &=& \sqrt{\sqrt{\frac{5}{2}}-\frac{1}{2}} \; t_{pd} 
	\sim 1.04\, t_{pd} , \label{mu_c1} \\
	\Delta_c  &=& 2\sqrt{\sqrt{\frac{5}{2}}-\frac{1}{2}} \; t_{pd} 
	\sim 2.08\, t_{pd}. \label{delta_c1}
\end{eqnarray}
Substituting eqs.(\ref{mu_c1}) and (\ref{delta_c1}) into eqs.(\ref{alphaD}), (\ref{alphaU}) and (\ref{ratio}), we obtain 
\begin{eqnarray}
C_{\Delta}^d &=& \frac{50(\sqrt{10}-1)}{15+29\sqrt{10}} \sim 1.01 \; ,\nn\\
C_U^d &=& \frac{\sqrt{5(\sqrt{10}-1)}(5+4\sqrt{10})}{58+3\sqrt{10}}
      \frac{t_{pd}}{U} \sim 0.860\frac{t_{pd}}{U}  ,\label{au}  \nn\\
\frac{C_{\Delta}^p}{C_{\Delta}^d} &=&\frac{C_{U}^p}{C_{U}^d} = 
      \sqrt{\frac{2}{5}}-\frac{1}{5} \sim 0.432 ,  \nn
\end{eqnarray}
in the limit $U\rightarrow\infty$. 
For $\Delta>\Delta_c$, we also solve eq.(\ref{Uinfty}) to obtain $\mu_\pm$, which yield the discontinuity in the chemical potential ${\it \Delta}\mu=\mu_+-\mu_-$ in the limit $U\rightarrow\infty$: 
\begin{eqnarray}
{\it \Delta}\mu =\sqrt{\Delta^2-\Delta_c^2}. 
\end{eqnarray}


In the limit $\Delta \rightarrow \infty$, eqs. (\ref{A1}) and (\ref{A2}) are simply written as 
\begin{eqnarray}
&&\hspace*{-7mm}	2t_{pd}^4(5U^2-2U\mu+2\mu^2)  \nn\\
	&&\hspace{10mm}-\{t_{pd}^2(U-2\mu)+2(U-\mu)\mu\Delta\}^2 = 0 ,\label{Deltainfty} \label{critical} \\
&&\hspace*{-7mm}	9t_{pd}^2 U^2 - \Delta\left(10U^3 - 22U^2\mu +6U\mu^2 - 4\mu^3\right)=0 , \label{critical2}
\end{eqnarray}
respectively. Here we assume 
\begin{eqnarray}
	\mu = \beta U \; , \label{mu_c}
\end{eqnarray}
and put eq.(\ref{mu_c}) into eq.(\ref{critical}), then we find
\begin{eqnarray}
\beta = \frac{1}{2}-\frac{9}{19}\gamma +O(\gamma^3),  \hspace{7mm}
\gamma = \frac{t_{pd}^2}{U\Delta }. \label{betagamma}
\end{eqnarray}
Using eq.(\ref{mu_c}) with (\ref{betagamma}) in eq.(\ref{critical2}), we have
\begin{eqnarray}
	\gamma &=& \frac{19}{3\sqrt{2(341+95\sqrt{13})}}  \sim 0.171 , \\
	\beta  &=& \frac{1}{2}-\frac{9}{19}\gamma \,\,\,\sim \,\,\,0.419 \; .
\end{eqnarray}
Then the critical values of the MIT are given by
\begin{eqnarray}
\mu_c&=& \frac{\beta}{\gamma}\frac{t_{pd}^2}{\Delta } \sim 2.45 \frac{t_{pd}^2}{\Delta }  , \label{mu_c2}\\ 
U_c&=&\frac{1}{\gamma}\frac{t_{pd}^2}{\Delta } \sim 5.84\,\frac{t_{pd}^2}{\Delta }, \label{U_c2}
\end{eqnarray}
in the limit $\Delta \rightarrow \infty$. Substituting eqs.(\ref{mu_c2}) and (\ref{U_c2}) into eqs. (\ref{alphaD}), (\ref{alphaU}) and (\ref{ratio}), we obtain
\begin{eqnarray}
C_{\Delta}^d &=& \frac{162 (3051853202 + 846431785 \sqrt{13})}
                 {348680453849 + 96706558085 \sqrt{13}} \sim 1.42, \nn\\
C_U^d &=&  \frac{15(16085875+4461419\sqrt{13})}
	            {114(1492608+413975\sqrt{13})}  \nn\\
      &-&  \frac{\sqrt{2}(880\sqrt{13}+3173)(341+95\sqrt{13})^{\frac12}}
	            {114(1492608+413975\sqrt{13})} \sim 1.42, \nn\\
\frac{C_{\Delta}^p}{C_{\Delta}^d} 
      &=&  \frac{C_{U}^p}{C_{U}^d}=
           \frac{104831 + 29075 \sqrt{13}}{18 (6508 + 1805\sqrt{13})}
           \frac{t_{pd}^2}{\Delta_c^2}\sim 0.895\frac{t_{pd}^2}{\Delta_c^2} \nn
\end{eqnarray}
in the limit $\Delta \rightarrow \infty$. For $\Delta=\infty$ and $U>U_c=0$, eq.(\ref{Deltainfty}) yields $\mu_+=U$ and $\mu_-=0$ resulting in 
\begin{eqnarray}
{\it \Delta}\mu=U.
\end{eqnarray}



\begin{thebibliography}{}
\bibitem{critical}
Critical behaviour of the other physical quantities and that for the doping dependence in the non-symmetric case were also discussed in ref. 23 
\bibitem{2band}
We can consider the more general case with the hopping integrals between $p$-$p$ and $d$-$d$ orbitals as discussed in ref. 22 
\bibitem{mu}
In the case with $n=1$, the chemical potential is restricted to the region $ \min(\Delta,0) < \mu < \min(\Delta,U)$. When we consider the MIT for $n=3$, it is restricted to the parameter region $\max(\Delta,0) < \mu < \max(\Delta,U)$. However, the MIT for $n=3$ is easily obtained from that for $n=1$ using the particle-hole transformation.
\end{thebibliography}
\end{document}